# Comparative analysis of the diffusion of Covid-19 infection in different countries

*Real-time analysis updated on March 16*

Fabio Miletto Granozio, CNR-SPIN, Napoli, IT
Email: fabio.miletto@spin.cnr.it

**Abstract**

The sudden spread of Covid-19 outside China has pushed on March 11 the World Health Organization to acknowledge the ongoing outbreak as a pandemic. It is crucial in this phase to understand what should countries which presently lag behind in the spread of the infection learn from countries where the infection spread earlier. The choice of this work is to prefer timeliness to comprehensiveness. By adopting a purely empirical approach, we will limit ourselves to identifying different phases in the plots of different countries, based on their different functional behaviour, and to make a comparative analysis. The comparative analysis of the registered cases curves highlights remarkable similarities, especially among Western countries, together with some minor but crucial differences. We highlight how timeliness can largely reduce the size of the individual national outbreaks, ultimately limiting the final death toll. Our data suggest that Western governments have not unfortunately shown the capability to anticipate their decisions, based on the experience of countries hit earlier by the outbreak.

**Introduction**

Italy is presently the hardest hit country. Its death toll seems bound to rapidly overcome the Chinese case. Other Western countries follow the same route. It is crucial in this phase to understand what countries presently lagging behind in the spread of the infection can learn from countries where the infection spread earlier.

The first question we address is: which are the relevant numerical signatures to be monitored to check how effectively a country is acting, compared to other countries, in containing the infection?

Here we show that an answer to this question can be given not relying on specific epidemiological expertise, but based on a simple numerical analysis of public data available on the internet. It is widely acknowledged that comparison of curves from different countries is made difficult by the different ways the detection of the virus is addressed. In particular, the ratios of the tested population to the total population, among the countries addressed in this paper, range from about 5.000/million in the case of South Korea, to 1.400/million for Italy to less the 70/million in the case of USA[1]. Therefore, not only the real number of infected people might largely exceed the registered cases, but such ratio (registered/total cases) might change country by country.

This difference is partly mitigated by two observations:

- in the specific spirit of this work, we will compare countries in the same stage of the epidemic. Considering the delay of the outbreaks (Korea is grossly 10 days ahead of Italy and 20 days ahead of the States, as will be discussed later in detail) Italy is grossly following the same testing curve as Korea while the USA lag behind by less than one order of magnitude.

- we should assume that in advanced countries most symptomatic patients are counted among the registered cases within a few days. These are exactly the cases we mostly want to focus on. The hidden background of asymptomatic patients, though playing a role in determining the disease spread, is a less relevant datum in foreseeing the final death count.

At the end of this analysis, remarkable similarities are found, especially among Western countries, together with some minor differences. The extent and relevance of the observed similarities for the case of Western countries will justify, ex post, the adopted approach.

Epidemiological curves are typically believed to follow the stochastic logistic model. Nevertheless, this assumption cannot take into account situations in which the infected population reacts by drastic changes of its collective behaviour, thus changing the virus reproductive number, in the course of the outbreak. A general model for Covid-19 diffusion would require knowledge of all the specific virus containment measurements adopted in each single country, their dates, and their quantitative effect on the reproductive number. This is so far beyond our present understanding.

The choice of this work is to prefer timeliness to comprehensiveness. By adopting a purely empirical approach, we will limit ourselves to identifying different phases in the plots of different countries, based on their different functional behaviour, and to make a comparative analysis. For more complex approaches, the human data analysis time could easily exceed the obsolescence time of the dataset, which is of the order of a couple of days.

**The case of countries beyond the initial phase of the outbreak**

The source of data for this work is the CSSE COVID-19 Dataset[2]. We analyse here the data of three of the countries that registered at the date of March 15 the highest cumulative number of registered cases, i.e. China, Italy, and South Korea. We neglect Iran, also because of the lack of information about the number of tested people. The three countries are at different stages of the outbreak. China exceeded the number of 1000 registered cases in the Hubei province on January 24 and presently reports few new cases per day. Korea exceeded the number of 1000 registered cases on February 27 and presently reports few new cases per day. Italy exceeded the number of 1000 registered cases only a few days later but, in spite of a recent slowdown, the end of the exponential phase, if confirmed, is happening in the present days.

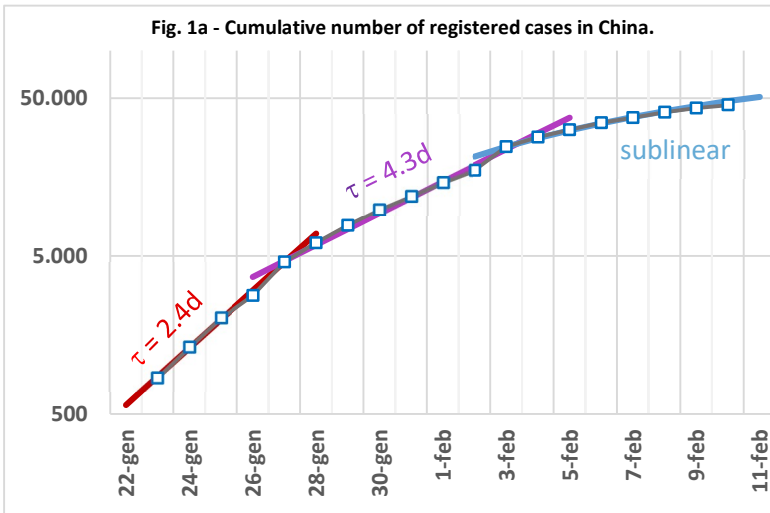

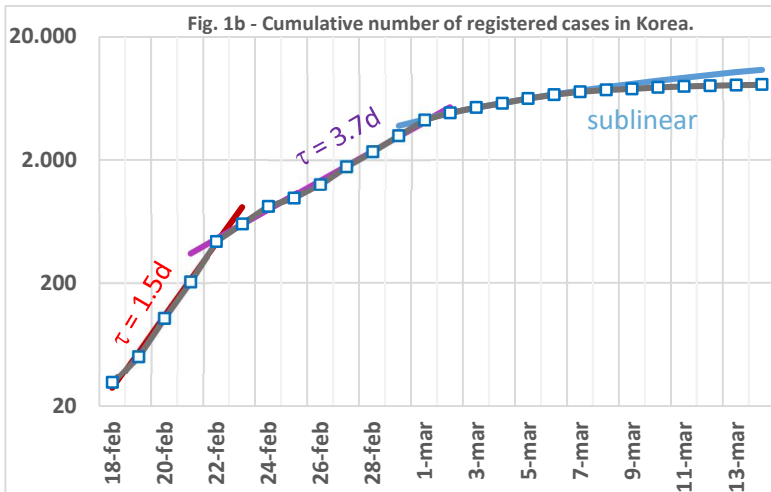

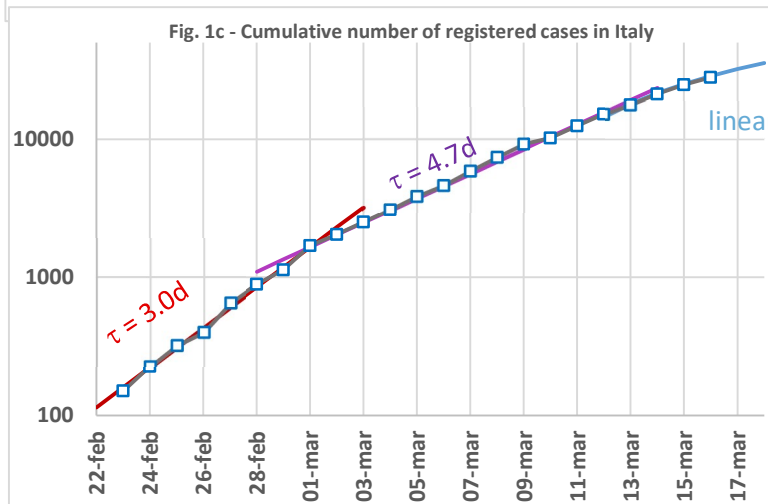

The Chinese plot shown in Fig. 1a stops on February 11. This is due to the change of criterium adopted in the counting of infected patients performed in China on February 12 [3]. The reported data are fortunately widely sufficient for the purpose of this work.

The plots of the three countries show similarities and differences. We observe a fast-growing exponential phase (#1) with a large country-to-country spread of the time constants $\tau$, here expressed in units of days (d) (CN: $\tau$ =2.4d; KR: $\tau$ =1.7d; IT: $\tau$ =3.0d). At some point, a pretty clear transition to a slower exponential growth (phase #2) takes place in each of the countries. The three exponents (CN: $\tau$ =4.3d; KR: $\tau$ =3.7d; IT: $\tau$ =4.7d) are close within ± 15%. After an interval of 8-10 days, the two Asian countries show another sudden switch to a quite different functional behaviour (phase #3), characterised by a sublinear growth, as clear shown by the comparison with the blue curve, representing linear growth (i.e., a straight line in linear scale). Such a switch allowed a remarkable slowdown in the growth of registered cases, preventing the infections from becoming unmanageably high.

As for Italy, last data are aligning to a linear curve, which might well be the inflection point, anticipating a smooth transition to a sublinear behaviour. The absolute number of the new daily cases in the country, about 3.500 on March 15, is still very high. By comparing the linear coefficient of the Italian and Korean blue curves, we observe in fact that the former exceeds the later by a factor 7.

The comparison of the plots shows that, in spite of the extremely fast growth rate ($\tau$ =2.4d, corresponding to a doubling time of one day) the rapid response of the Korean society allowed to switch the growth to a slower rate before reaching 500 registered infected people. This rapidity is confirmed by the observation that at the time when only 21 infections were found, on February 18, the Republic of Korea had already tested over 8000 citizens[4]. Italy had instead 322 registered cases by the time it reached the same number of tests on February 25[5]. The perduring fast growth rate in Italy rises major concerns and suggests that the cumulative final number of cases might exceed the Chinese case.

**Comparison between Western countries.**

The Italian case seems to correspond to the worst-case scenario among the ones analysed above. It suggests that while Korea implemented a faster reaction than China, profiting of the lesson learned by the experience of the neighbour country, Italy apparently showed a longer response time than both Asian countries, in terms either of diagnosis, or of governmental decision, or else of change of individual habits. It is of great importance to verify how well the Italian lesson was learned by other Western countries.

The plots reported in Fig. 2a,b,c,d show the curves of the four Western countries reporting the highest amounts of registered cases: Spain, Germany, France and United States. For each of them is it observed that, excluding a very initial phase, in which numbers were small and probably affected by the insufficient tests number, all datasets align to exponential red curves with exponent $\tau$ ~2.9-3.0d, corresponding to a doubling time $\tau_D$ ~2.0d.

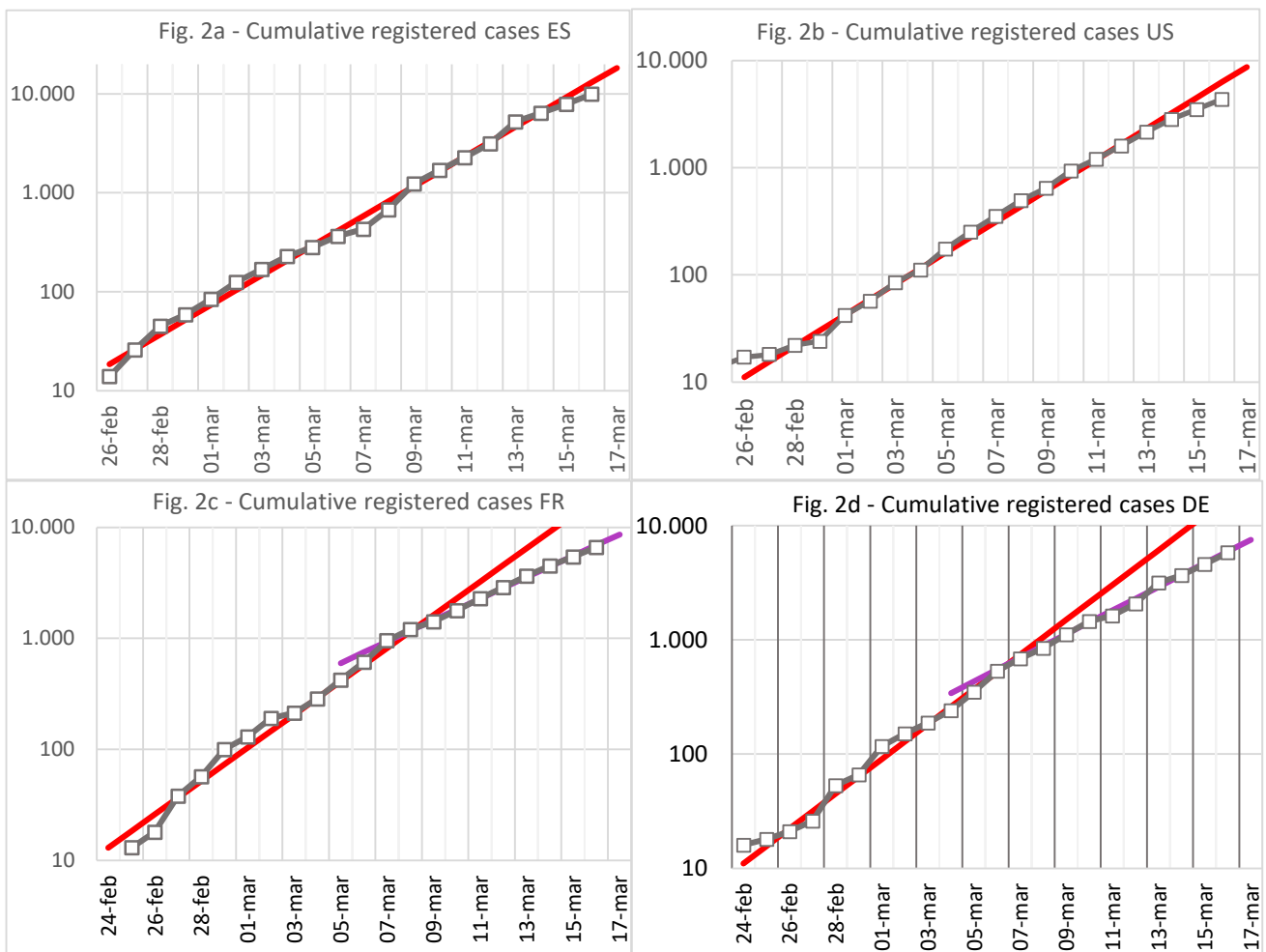

This timescale seems to be characteristic of the Covid-19 "free expansion" in all these countries or, to state it more prudently, of the rate at which they are detected. We remind in this context that, in absence of large-scale screening programs, most infected people are tested after showing symptoms, i.e. about 5 to 14 days after infection. Therefore, the registered cases curves map the history of the past behaviours of the infected population.

We observe that, still at the date of March 15, the Spanish evolution is correctly described by the red exponential curve. The US curve shows a minor deviation, which might well indicate a switch to phase #2. France has switched to the second phase, with exponent $\tau$ =4.5d. Germany has also switched to the second phase, with exponent $\tau$ =4.2d.

The plot in Fig. 3 gathers all the curves above in the same plot, comparing them with the Italian curve. Among the possible ways to plot the data together, the most significant one, by far, was found to apply a relative shift in time, in order to "synchronize" the different starting times of the outbreak. We remark that, by normalizing the number of infected people to the overall population, the US plot would have been shifted backwards by 5-6 days.

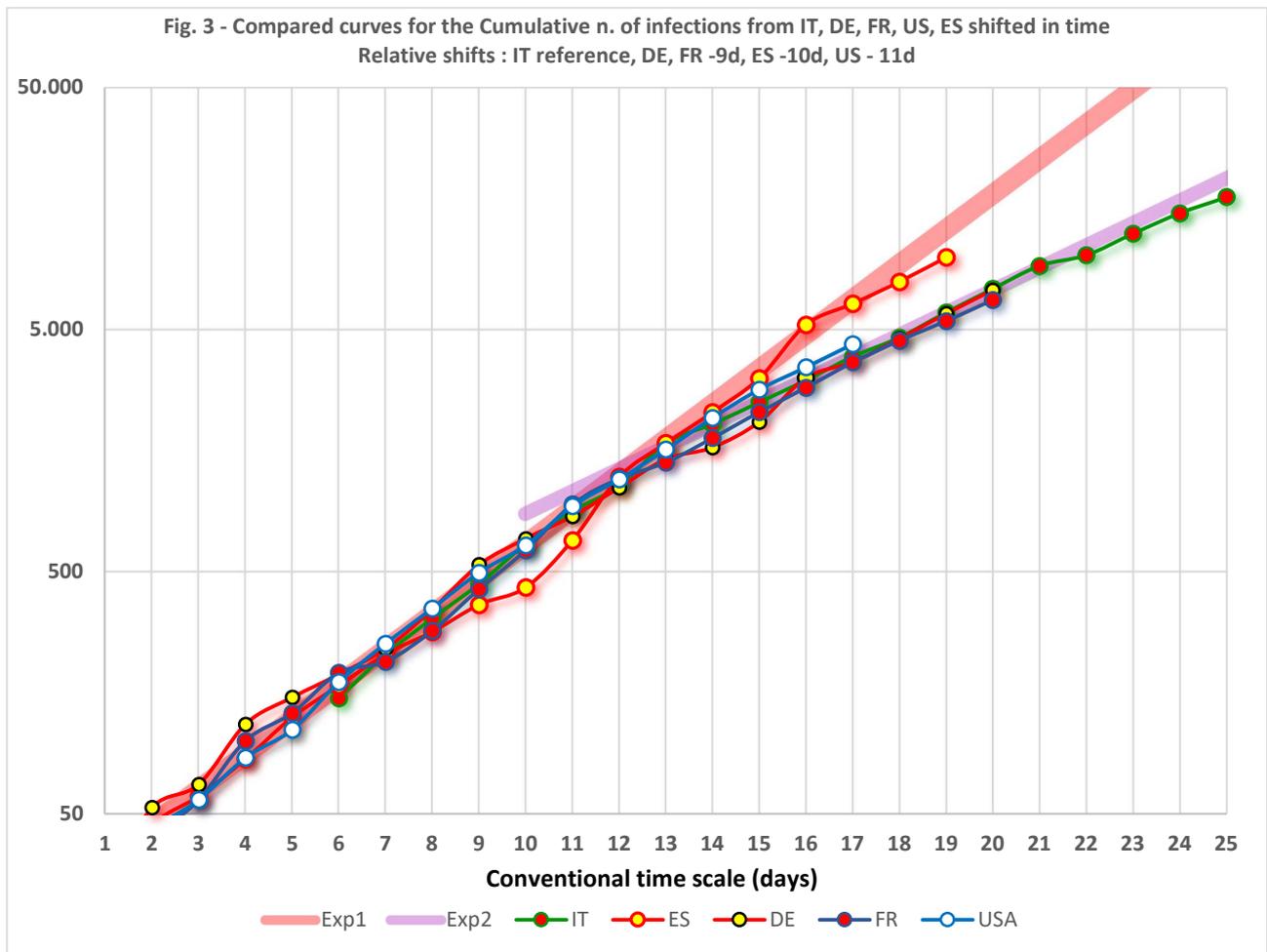

Fig. 3 - Compared curves for the Cumulative n. of infections from IT, DE, FR, US, ES shifted in time
Relative shifts : IT reference, DE, FR -9d, ES -10d, US - 11d

When plotted with the appropriate relative time scale (IT reference, DE, FR -9d, ES -10d, US - 11d), the data show how early or late the different countries deviated from the red exponential "phase #1" curve with $\tau$ ~2.0d, $\tau_D$ ~2.0d. The violet curve fitting the Italian "phase #2" is also shown. It can be qualitatively deduced that France, Germany and probably the United States, on this particular conventional time scale, have switched to the violet curve grossly at the same time as Italy did. Spain is potentially running towards a worse scenario, although the last points hint to a possible alignment to the same violet curve, albeit a factor two above.

## Importance of timely response

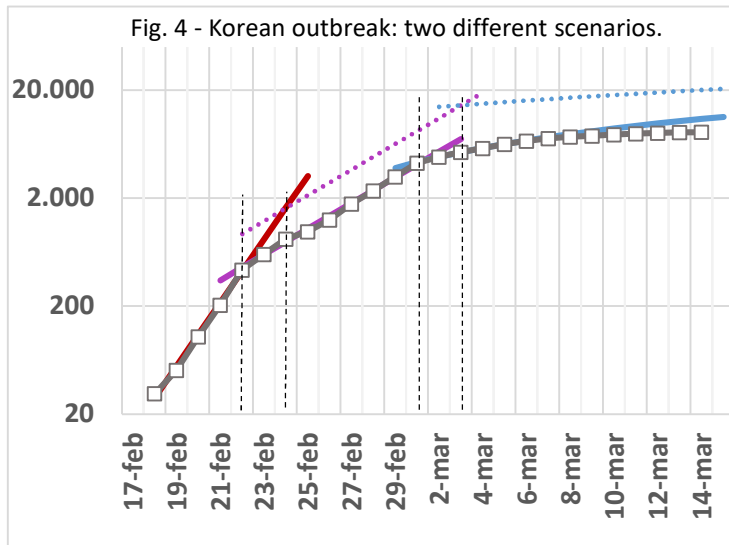

Fig. 4 - Korean outbreak: two different scenarios.

The plot in Fig. 4 graphically highlights the importance of the early reactions. The numerical history of the outbreak in Korea is compared to the hypothetical outbreak evolution (dotted lines) in case a two-day delay in the transition from the red to the violet curve, rigidly reflected in a two-day shift in the transition to the blue curve, would have taken place.

According to the estimation in Fig. 4, the transition to phase #3 would have taken place on March 3 with an infected population of about 14.500 people, about 3,5 times higher than the actual number of registered cases at the real transition, happened on March 1st. The same scale factor would be applied today, within our hypotheses, to the actual infected population.

## Conclusion

We attempted an elementary, real-time analysis of the Covid-19 diffusion data updated at March 16. Timeliness was preferred to comprehensiveness. Important information has been extracted by the data, but major caution is needed in deriving general and far-reaching conclusions. Both the inhomogeneity in data acquisition rate in different countries and the huge background of undetected, presumably asymptomatic, infected patients, are two major sources of uncertainty. The criterion applied in the plot in Fig. 3 is highly instructive but, to some extent, arbitrary.

With all due prudence related to the uncertainties above, we believe the present analysis is an excellent and timely starting point for further studies on the delayed effects on the curves of the response adopted by different countries. Such response includes both individual changes of habits (hands hygiene, social distancing by own choice) and restrictions imposed by the governments (closing of schools, constraints to the mobility of citizens).

The Korean example clearly shows that early diagnosis of the first infected patients and timeliness in the response can largely reduce the size of the outbreak, ultimately limiting the final death toll. Our data suggest that Western governments have not shown the capability to anticipate their decisions based on the experience of countries hit earlier by the outbreak. Our hope is that this work can contribute to triggering early and appropriate responses to the Covid-19 pandemic.

## References

1. How many tests for COVID-19 are being performed around the world? Our World in Data. https://ourworldindata.org/covid-testing-10march. Accessed March 12, 2020.
2. CSSEGISandData. *CSSEGISandData/COVID-19*.; 2020. https://github.com/CSSEGISandData/COVID-19. Accessed March 15, 2020.
3. Coronavirus Cases: Statistics and Charts - Worldometer. https://www.worldometers.info/coronavirus/coronavirus-cases/#cases-growth-factor. Accessed March 10, 2020.
4. https://www.cdc.go.kr/board/board.es?mid=&bid=0030. Accessed March 15, 2020.
5. Quanti test per il coronavirus abbiamo fatto. Il Post. http://www.ilpost.it/2020/02/25/tamponi-coronavirus-italia-regno-unito-francia/. Published February 25, 2020. Accessed March 15, 2020.